# Compact all-fiber polarization-independent up-conversion single-photon detector

Long-Yue Liang,[a,b] Jun-Sheng Liang,[c] Quan Yao,[a] Ming-Yang Zheng,[a,c] Xiu-Ping Xie,[a,c] Hong Liu,[b] Qiang Zhang,[a,d,*] and Jian-Wei Pan[d]

[a] Jinan Institute of Quantum Technology, Jinan, 250101, China
[b] Shandong University, State Key Laboratory of Crystal Materials, Jinan, 250100, China
[c] Shandong Institute of Quantum Science and Technology Co., Ltd., Jinan, 250101, China
[d] University of Science and Technology of China, National Laboratory for Physical Sciences at Microscale and Department of Modern Physics, Hefei, 230026, China
[*] Corresponding authors. *E-mail addresses:* qiangzh@ustc.edu.cn (Q. Zhang)

**Abstract.**

We demonstrate a compact all-fiber polarization-independent up-conversion single-photon detector based on integrated reverse proton exchanged periodically poled lithium niobate waveguides. The horizontally and vertically polarized components of randomly polarized signals are separated with a fiber-coupled polarization beam splitter, launched into two orthogonally polarized polarization maintaining fibers and fetched into two adjacent independent waveguides on the same device. The up-converted outputs from both waveguide channels are then combined with a multi-mode fiber combiner and detected by a silicon detector. With this configuration, the polarization-independent single-photon counting at 1.55 μm is achieved with a system detection efficiency of 29.3%, a dark count rate of 1600 counts per second, and a polarization dependent loss of 0.1dB. This compact all-fiber system is robust and has great application potential in practical quantum key distribution systems.

Keywords：polarization-independent; single-photon detector; up-conversion; periodically poled lithium niobate waveguide.

## 1. Introduction

Telecom band single-photon detectors (SPDs), including InGaAs/InP avalanche photodiodes (APDs) [1,2], superconducting SPDs [3,4], and up-conversion SPDs [5,6], have been rapidly developed due to their extensive applications in optical quantum information [7], optical time domain reflectometry (OTDR) [8,9], single-photon-level spectrometer [10,11], laser detection and ranging (LADAR) [12], and other important applications.

Commercial InGaAs/InP SPDs are currently suffering from low detection efficiency of 10% corresponding to a dark count rate of 1000 counts per second (cps) or large dark count rate of 5000 cps corresponding to a detection efficiency of 20% [13]. Despite of their impressive performance, superconducting SPDs are limited in field implementation due to the need of bulky cryogenic cooling and exquisite temperature control [14,15]. On the other hand, the performance of up-conversion SPD satisfies the need of long-distance quantum key distribution (QKD) and is suitable for field implementation [16,17], where the detection spectral range of silicon APDs is extended into the 1.55-µm telecom band using sum-frequency generation (SFG) with periodically poled lithium niobate (PPLN) waveguides [6].

As a low noise and high-efficiency wavelength conversion device, the proton exchanged PPLN waveguide guides only the light wave polarized along the c-axis of the z-cut crystal [18], and all the interacting waves are of the same polarization for type-0 SFG, making use of the highest nonlinear coefficient $d_{33}$. Up-conversion SPD based on a single waveguide is therefore polarization dependent and extra caution should be taken when up-conversion SPD is used in polarization-independent systems, such as time-bin phase-encoding QKD system which has been proved to be a practical scheme [19-21] in complex environmental conditions where the polarization states change randomly during long distance fiber transmission [22-24], quantum lidar [25], single photon imaging [26], and biomedical luminescence spectroscopy [27] for which the detected photon polarization is also random.

A method to realize polarization-independent up-conversion SPD had been proposed by Takesue et al [28], which used a polarization diversity configuration composed of two up-conversion SPDs, dedicated for the two orthogonal polarization components respectively. The system detection efficiency was 2.3% at 1550 nm with a dark count rate of 14 000 cps. The performance of this detector limits its applications in many fields. As is known to us all, titanium in-diffused PPLN waveguides could support the propagation of both horizontal (H) and vertical (V) polarization components [29]. However, the different quasi-phase-matching (QPM) conditions for H and V polarization components caused by refractive index discrepancy determined that only one component of the randomly polarized signal could be up-converted for a single waveguide [30,31].

In this paper, we demonstrate a compact all-fiber polarization-independent up-conversion detector for single photon counting at 1.55 μm based on the polarization diversity configuration. An integrated dual-channel reverse proton exchanged (RPE) PPLN waveguide device with two adjacent waveguides of the same design parameters is used. After being separated with a fiber-coupled polarization beam splitter (PBS), the H and the V polarized components of the randomly polarized signal pass two orthogonally polarized polarization maintaining (PM) fibers, combine with the pump waves and enter the waveguide device. Optical filtering for the outputs is achieved by combining multi-mode fiber filter with multi-mode fiber combiner (MMFC). Such an all-fiber configuration makes the system compact and stable. Based on this detector, efficient polarization-independent single-photon counting at 1.55 μm is demonstrated, showing a system detection efficiency of 29.3%, a dark count rate of 1600 cps and a polarization dependent loss of 0.1dB. The detection efficiency remains stable when the polarization of the signal is changed, and this detector can be used in practical QKD systems for complex environmental applications.

## 2. Experimental Setup and Waveguide Characterization

*2.1. Experimental Setup*

Fig. 1 shows the schematic diagram of the polarization-independent up-conversion SPD.

As shown in Fig. 1(a), a single-frequency fiber laser fixed at 1950 nm serves as the pump source. The 1550 nm light wave from a TSL-510 tunable semiconductor laser is attenuated to a single-photon level of $10^6$ photons per second using two 1550-nm variable optical attenuators (VOAs) and a 1/99 1550-nm PM beam splitter (BS) and serves as the signal source. A calibrated power meter is exploited to monitor the input signal power [32]. We choose the commercial 1950-nm fiber laser as the SFG pump based on that the long-wavelength pumping technique can dramatically reduce the noise photons generated by the spontaneous scattering processes of the strong pump [33]. Such a design achieves much lower noise-equivalent-power than the short-wavelength pumping technique [34] at the expense of missing the highest detection efficiency of commercial silicon APD at around 700 nm.

As shown in Fig. 1(b), the 1550-nm signal is launched into a fiber-coupled PBS and projected to the two linear polarization states of H and V. Each polarization component is combined with one output port of a dual-channel 1950-ncm pump laser in a wavelength division multiplexing (WDM) coupler and then pigtailed into the dual-channel PPLN waveguide device. Two 1950-nm VOAs are inserted to adjust the required pump power for the waveguides. PM fibers are used for all the connections from light sources to the waveguide device, and the fibers between the PBS and WDMs are appropriately twisted to ensure that the signal is vertically polarized before entering the waveguides, minimizing coupling losses for the signal. The up-conversion photons generated from the two SFG waveguide channels are combined with a MMFC, therefore only one set of fiber filter and silicon APD is needed. To filter the noise photons generated from the strong pump, the fiber filter consists of an 857-nm band-pass filter with a bandwidth of 30 nm and an 863.4-nm narrow-band filter with a bandwidth of 1.2 nm. The noise photons mainly come from the spontaneous Raman scattering noise, parasitic noise caused by imperfect periodic poling structures, and second and third harmonic generation of the pump [33]. Finally, the SFG photons are detected by a silicon APD with a detection efficiency of 55% in 850-nm band and a dark count rate of 200 cps.

The working temperature of the PPLN waveguide is kept at 37.5 °C by thermoelectric cooling to maintain the phase-matching condition. A polarization controller (PC) is used to change the polarization states of the signal photons.

*2.2. Waveguide Characterization*

We fabricate the PPLN waveguides with the RPE technique [35]. The dual-channel PPLN waveguide device is composed of two adjacent independent waveguides with the same design parameters and a center-to-center separation of 127 μm. A PM fiber array consisting of two 1550-nm PM fibers terminated in a V-groove structure with a core spacing of 127 μm is used for input coupling. A 1-mm-long 4.0-μm-wide mode filter is located at the input port of the waveguides to match the mode size of 1550-nm PM fibers, followed by a 1-mm-long linear taper increasing the waveguide widths from 4.0 μm to 8.0 μm, with the latter kept through the remaining waveguide. The total length of the PPLN waveguide device is 52 mm, including 48-mm-long QPM gratings with a period of 20 μm. A multi-mode fiber array is used for output coupling. Both the input and output end facets of the PPLN waveguide are anti-reflection coated to eliminate the Fresnel reflection loss.

The phase-matching wavelengths and conversion efficiencies of the two up-conversion waveguides are measured respectively.

As shown in Fig. 2(a), the SFG tuning curves of the two adjacent waveguides (marked as waveguide 1 and 2) in the dual-channel PPLN waveguide device are obtained by sweeping the signal wavelength around 1550 nm with the pump wavelength fixed at 1950 nm. The phase-matching wavelengths of waveguide 1 and 2 are 1546.58 nm and 1546.62 nm at 25 °C with the full width at half maximum (FWHM) of 0.61 nm and 0.62 nm respectively. Consistency of the phase-matching wavelengths for two adjacent independent waveguides on a single chip enables high conversion efficiencies for H and V polarizations simultaneously.

Fig. 2(b) shows the conversion efficiencies versus the pump power measured at the output port of the waveguides. As illustrated, the signal photon conversion efficiency of waveguide 1 reaches its maximum when the pump power is 115 mW, corresponding to a normalized conversion efficiency of 93.1%/(W cm$^2$), which is calculated with [34]

$$\eta_{nor} = \frac{\pi^2}{4L^2 P_{max}}, \qquad (1)$$

where $\eta_{nor}$ is the normalized conversion efficiency, $L$ is the length of the QPM gratings, and $P_{max}$ is pump power required for maximum conversion. While for waveguide 2, the maximum conversion efficiency is obtained when the pump power is 110 mW, corresponding to a normalized conversion efficiency of 97.3%/(W cm$^2$). This difference may come from fabrication errors including the random duty-cycle errors of the QPM gratings and the waveguide width errors.

The signal photon conversion efficiency η is

$$\eta = \frac{P_{SFG} \cdot \lambda_{SFG}}{P_{signal} \cdot \lambda_{Signal}}, \qquad (2)$$

where $P_{SFG}$ is the SFG power at the output port of the waveguide, $P_{Signal}$ is the signal power at the input port of the waveguide, and $\lambda_{SFG}$ and $\lambda_{Signal}$ are the SFG and signal wavelengths. Calculated from data taken at the maximum points of the curves in Fig. 2(b), the maximum signal photon conversion efficiencies are 73.8% for waveguide 1 and 74.3% for waveguide 2 respectively.

## 3. Polarization-independent Up-conversion Single-photon Detection

To obtain a polarization-independent up-conversion SPD, the detection efficiencies for H and V polarizations should be equal. By selecting appropriate components and adjusting the pump power using 1950-nm VOA to compensate for the differences of the dual-channel pump power and maximum signal photon conversion efficiencies between the two waveguides, the overall maximum detection efficiencies of both waveguide channels are adjusted to the same value of 54.0%. Therefore the overall detection efficiencies of the two waveguide channels are the same for any pump power out of the laser. The throughputs of the components in the polarization-independent up-conversion SPD are summarized in Table 1.

For randomly polarized signal input, we record the detection efficiency and dark count rate of the polarization-independent up-conversion SPD by gradually increasing the pump power out of laser. The detection efficiency is calculated by dividing the detected count rate (after subtracting the dark count rate) by $10^6$ (the total input of signal photon), and the dark count rate contains the intrinsic noise of the silicon APD. As shown in Fig. 3, the maximum detection efficiency is 29.3% with a dark count rate of 1600 cps, matching well with the theoretical detection efficiency of 29.7% by combining the throughputs of the components (54%, from Table 1) and the detection efficiency of silicon APD (55%).

We further confirm that the detection efficiency of this polarization-independent up-conversion SPD remains stable when the polarization of the signal is changed. Fig. 4 shows the detected count rates versus the proportion of V polarization component, where the black hollow squares, red hollow circles, and blue solid triangles denote the count rates for V polarization, H polarization, and the signal polarization respectively. For each data point, the proportion of V polarization component of the signal is measured at the output ports of the PBS with a signal power of several milliwatts. Then the signal power is attenuated to a single-photon level of $10^6$ photons per second at the input port of the PBS and the corresponding count rate is recorded. The non-polarizable devices, including the VOAs, 1/99 BS, and PC, are all kept stable to maintain the input polarization unchanged for each measurement. As illustrated, with the increase of the proportion of V polarization component, the count rate for V polarization increases and that for H polarization decreases, while the total count rate remains stable. The system detection efficiency is about 29.3% after subtracting the dark count rate when the signal polarization is tuned.

The polarization dependent loss (PDL) of our detector is [36]

$$\text{PDL} = 10\log\frac{\text{PIE}_{max}}{\text{PIE}_{min}}, \tag{3}$$

where *PIE*$_{max}$ and *PIE*$_{min}$ are the maximum and minimum detection efficiencies of the polarization-independent system for a randomly polarized signal input. Experimental data in Fig. 4 yields a PDL of 0.1 dB which is similar to common optical devices, such as optical isolator, dense wavelength division multiplexing (DWDM), etc.

When our dual-channel PPLN waveguide device is used as two polarization dependent up-conversion SPDs, the maximum detection efficiencies of the two waveguide channels are 35.2% and 36.4%, with dark count rates of 950 cps and 850 cps respectively. The polarization-independent up-conversion SPD has a slightly lower (-0.8 dB) maximum detection efficiency due to the usage of PBS and MMFC, while the dark count rate is almost doubled because the pump power is twice for using two independent waveguides. These tradeoffs are worthwhile when polarization independent measurement is of priority in the applications.

## 4. Conclusions

We demonstrate a compact all-fiber polarization-independent up-conversion SPD based on an integrated dual-channel PPLN waveguide device in the polarization diversity configuration. Based on this detector, the polarization-independent single-photon counting at 1.55 μm is demonstrated, achieving a high detection efficiency of 29.3%, a low dark count rate of 1600 cps and a polarization dependent loss of 0.1dB. These results imply that our detector is of great application potential in polarization-independent systems, such as time-bin phase-encoding QKD, quantum lidar, single photon imaging and biomedical luminescence spectroscopy. We may integrate the WDMs in the waveguide device by using directional couplers and further improve the compactness of the system.

**Disclosures**

Declarations of interest: none.

**Acknowledgements**

This work was supported by the National Key R&D Program of China [Grant Numbers 2018YFB0504300]; the Chinese Academy of Science; the SAICT Experts Program; the Taishan Scholar Program of Shandong Province; the Shandong Peninsula National Innovation Park

**Table 1.** Throughputs of the components in the polarization-independent up-conversion SPD

| Component | Throughput | |
|---|---|---|
| | Waveguide channel 1 | Waveguide channel 2 |
| PBS @1550 nm | 89% | |
| WDM @1550 nm | 91.8% | 94.3% |
| Waveguide (Signal photon conversion efficiency from 1550 nm to 863 nm) | 73.8% | 74.2% |
| MMFC @863 nm | 94.8% | 91.8% |
| Filter @863 nm | 94.5% | |
| Overall maximum detection efficiency | 54.0% | |

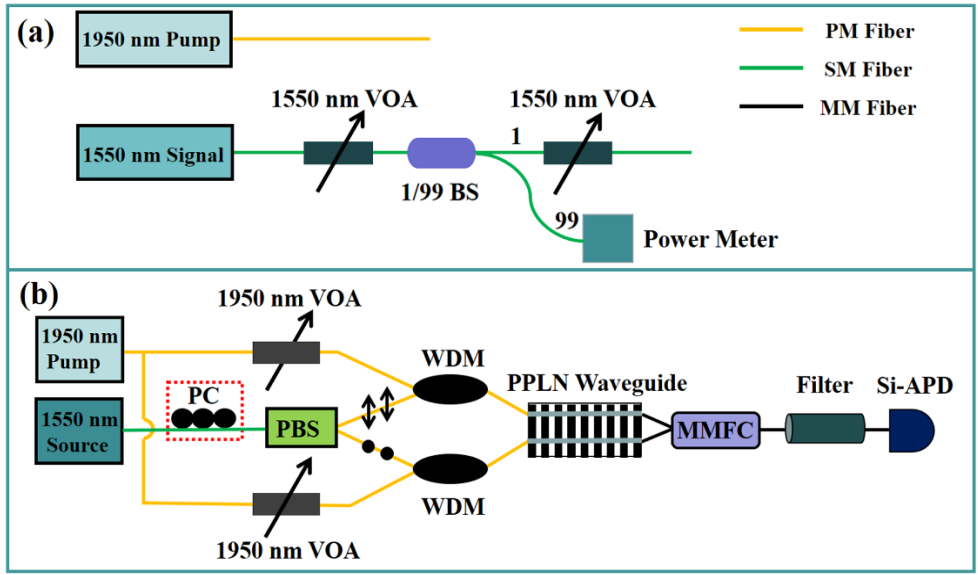

**Fig. 1** (a) Preparation of the pump and signal sources. (b) Schematic diagram of the polarization-independent up-conversion SPD. VOA: variable optical attenuator, BS: beam splitter, PM: polarization maintaining, SM: single-mode, MM: multi-mode, PC: polarization controller, PBS: polarization beam splitter, WDM: wavelength division multiplexer, PPLN: periodically poled lithium niobate, MMFC: multi-mode fiber combiner, Si-APD: silicon avalanche photodiode.

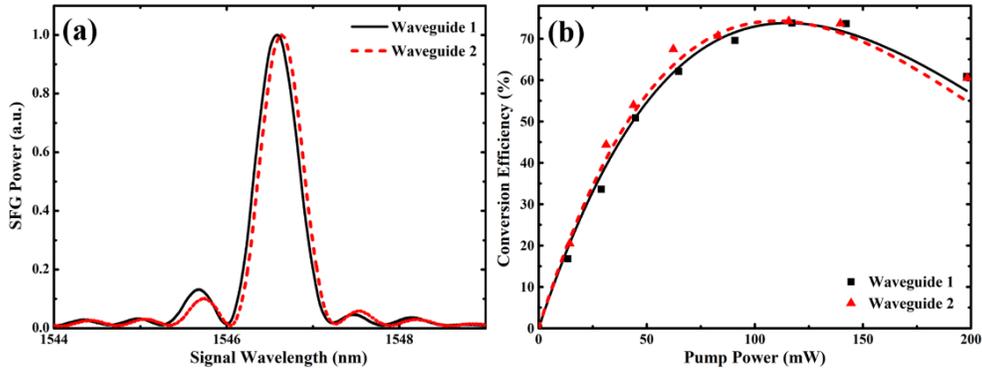

**Fig. 2** (a) SFG tuning curves, (b) Measured signal photon conversion efficiency (black square and red circle) and the $\sin^2()$ fitting results (black solid and red dashed lines) [33], the pump power is measured at the output port of the waveguides.

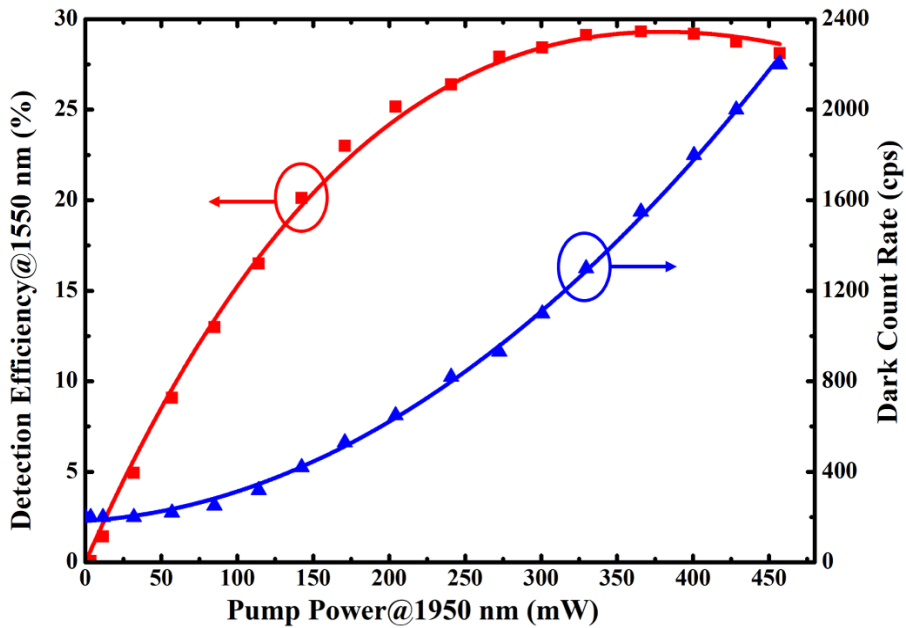

**Fig. 3** Measured detection efficiency (red square) and dark count rate (blue triangle) versus the pump power (sum of the input pump power for waveguide 1 and 2 at the input ports of the WDM couplers). The red and blue solid lines represent the fitting results.

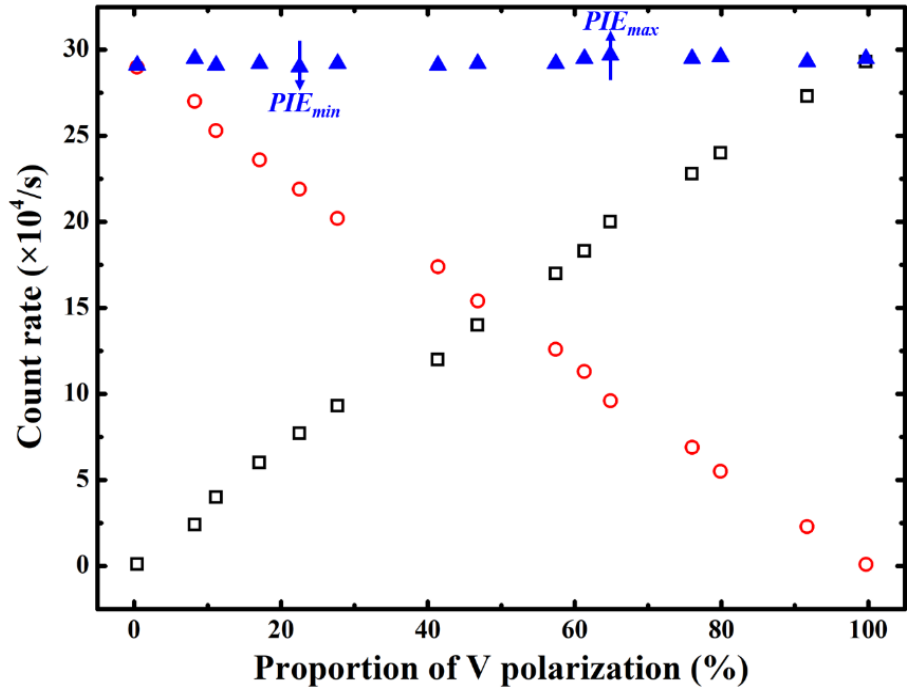

**Fig. 4** The count rates versus the proportion of V polarization component of the signal. Black hollow squares, red hollow circles, and blue solid triangles denote the count rates for V polarization, H polarization, and the signal polarization, respectively. *PIE$_{max}$* and *PIE$_{min}$* are the maximum and minimum detection efficiencies points when tuning the signal polarization.